 \documentclass{aa}  

\usepackage{graphicx}
\usepackage{txfonts}
\usepackage{subfig}             % handles subfloats
\usepackage{epstopdf}           % .eps to .pdf

\def\ltsima{$\; \buildrel < \over \sim \;$}
\def\simlt{\lower.5ex\hbox{\ltsima}}
\def\gtsima{$\; \buildrel > \over \sim \;$}
\def\simgt{\lower.5ex\hbox{\gtsima}}
\def\cgs{{erg cm$^{-2}$ s$^{-1}$}}
\def\ergs{{erg s$^{-1}$}}
\def\cm2{{cm$^{-2}$}}

\def\lum{{$L_{\rm 2-10}$}}

\def\p1{{Paper I}}

\def\xmm{{\em XMM--Newton}}
\def\chandra{{\em Chandra}}

\def\xmm{{\em XMM--Newton}}

\def\nh{{N$_{\rm H}$}}

\def\xray{{X--ray}}
\def\f14{{10$^{-14}$}}
\def\f13{{10$^{-13}$}}
\def\f12{{10$^{-12}$}}
\def\f11{{10$^{-11}$}}

\def\4u{{4U~1344$-$60}}

\def\feka{{Fe K$\alpha$}}

\def\lmir{{$L_{\rm 5.8\mu m}$}}

\def\lbol{{$L_{\rm Bol}$}}

\def\msun{{$M_{\rm \odot}$}}

\def\nus{{\em NuSTAR}}

\usepackage{color}
\usepackage{amstext}

\begin{document}

\title{The most obscured AGN in the COSMOS field}
           
   %\subtitle{I. Overviewing the $\kappa$-mechanism}

   \author{G. Lanzuisi\inst{1,2}
   \and
   M. Perna\inst{1,2}
   \and
   I. Delvecchio\inst{1,3}
   \and
   S. Berta\inst{4}
   \and
   M. Brusa\inst{1,2}
   \and
   N. Cappelluti\inst{2}
   \and
   A. Comastri\inst{2}
   \and
   R. Gilli\inst{2}
   \and
   C. Gruppioni\inst{2}
   \and
   M. Mignoli\inst{2}
   \and
   F. Pozzi\inst{1,2}
   \and
   G. Vietri\inst{5}
   \and
   C. Vignali\inst{1,2}
   \and
   G. Zamorani\inst{2}.
   }

 \titlerunning{The most obscured AGN in the COSMOS field}\authorrunning{G.~Lanzuisi et al.}

%\email{}

\institute{Dipartimento di Fisica e Astronomia, Universit\`a  di Bologna, Viale Berti Pichat 6/2, I-40127 Bologna, Italy \and
INAF - Osservatorio Astronomico di Bologna,  Via Ranzani 1, I--40127 Bologna, Italy \and 
Physics Department, University of Zagreb, Bijenicka cesta 32, 10002 Zagreb, Croatia \and 
Max-Planck-Institut f\"ur extraterrestrische Physik,  Giessenbachstrasse, 85748 Garching, Germany \and
INAF - Osservatorio Astronomico di Roma, via Frascati 33, 00040, Monte Porzio Catone (RM), Italy 
}
   \date{Received March 6 2015; Accepted April 17 2015}

  \abstract
{Highly obscured active galactic nuclei (AGN) are common in nearby galaxies, but are difficult to observe beyond the local Universe, 
where they are expected to significantly contribute to the black hole accretion rate density. 
Furthermore, Compton-thick (CT) absorbers (\nh\simgt10$^{24}$ \cm2) suppress even the hard X-ray (2-10 keV) AGN nuclear emission, 
and therefore the column density distribution above 10$^{24}$ \cm2 is largely unknown.
We present the identification and multi-wavelength properties of a heavily obscured (\nh$\simgt10^{25}$ \cm2), 
intrinsically luminous (\lum$>10^{44}$ \ergs) AGN
at $z=0.353$ in the COSMOS field. Several independent indicators, such as the shape of the X-ray spectrum, 
the decomposition of the spectral energy distribution and X-ray/[NeV] and X-ray/$6\mu m$ luminosity ratios, agree  
on the fact that the nuclear emission must be suppressed by a $\simgt10^{25}$ \cm2 column density.
The host galaxy properties show that this highly obscured AGN is hosted in a massive star-forming galaxy, 
showing a barred morphology, which is known to correlate with the presence of CT absorbers.
Finally, asymmetric and blueshifted components in several optical high-ionization emission lines
indicate the presence of a galactic outflow, possibly driven by the intense AGN activity (\lbol/$L_{Edd} = 0.3-0.5$). 
Such highly obscured, highly accreting AGN are intrinsically very rare at low redshift, whereas
they are expected to be much more common at the peak of the star
formation and BH accretion history,
at $z\sim2-3$. We demonstrate that a fully multi-wavelength approach can recover a sizable sample  
of such peculiar sources in large and deep surveys such as COSMOS.}

 \keywords{Galaxies:~individual -- Galaxies:~active --  Galaxies:~nuclei -- X-ray:~galaxies }
   
   \maketitle
%
%________________________________________________________________

\section{Introduction}

Highly obscured, Compton thick (CT, column density \nh\simgt10$^{24}$ \cm2) AGN are common in the local Universe, 
representing $\sim20-30\%$ of the local AGN population (Maiolino et al. 1998; Risaliti et al. 1999; Burlon et al. 2011).
Several AGN-galaxy co-evolutionary models propose that their high-redshift counterparts trace a crucial step in the AGN 
and galaxy build-up, corresponding to the most efficient phase of black hoe (BH) growth and star formation activity 
(e.g. Hopkins et al. 2008), possibly, but not always, triggered by major mergers (Draper \& Ballantyne 2012).
A large population of CT AGN is also required to reproduce the shape of the cosmic X-ray background (Comastri et al. 1995; Gilli et al. 2007). 
More recently, the existence of a large population of heavily CT (HCT, \nh$=10^{25-26}$ \cm2) 
AGN has been proposed (Comastri et al. 2015) to reconcile the increased normalization of the local super massive black hole (SMBH) mass density 
(Graham \& Scott 2013; Kormendy \& Ho 2013) with the normalization expected from growing black holes integrated over the cosmic time assuming
an accretion efficiency of  $\epsilon\sim0.1$.
Identifying these extremely obscured sources is, by definition, very difficult, even in the local Universe.
An example of this is represented by Arp220: after years of debate (Genzel et al. 1998; Iwasawa et al. 2001; Clements et al. 2002; Nardini et al. 2010), 
ALMA observations have finally shown that this prototypical hyper-luminous infrared (IR) galaxy indeed hosts an AGN with 
an estimated column density of \nh=$0.6-1.8\times10^{25}$ \cm2 (Wilson et al. 2014; Scoville et al. 2015).

The identification of CT AGN beyond the local Universe is even more challenging.
Recent results on preliminary \nus\ data (Harrison et al. 2013) were only able to place an upper limit ($< 33\%$) 
on the fraction of CT quasar (L$_{10-40keV}>10^{44}$ \ergs) between $z = 0.5 - 1$ (Alexander et al. 2013).
Moreover, given the lack of large, complete samples of distant AGN observed at $20-30$ keV, 
the relative fraction of HCT and CT sources remains poorly constrained and usually is not explored at all 
in deep survey studies (Tozzi et al. 2006; Ueda et al. 2014; but see also Brightman et al. 2014 and Buchner et al. 2015). 

In this article we present the discovery of a bona-fide HCT candidate at $z=0.353$ in the COSMOS field (Scoville et al. 2007), 
together with its multi-wavelength properties.
Throughout this paper we assume  $H_0$ = 70 km s$^{-1}$ Mpc$^{-1}$, $\Omega_\Lambda$ = 0.73, and  $\Omega_M$ = 0.27,
and a Chabrier (2003) initial mass function.

%======================================================
 \begin{figure*}[t]
 \begin{center}
 \includegraphics[width=7cm]{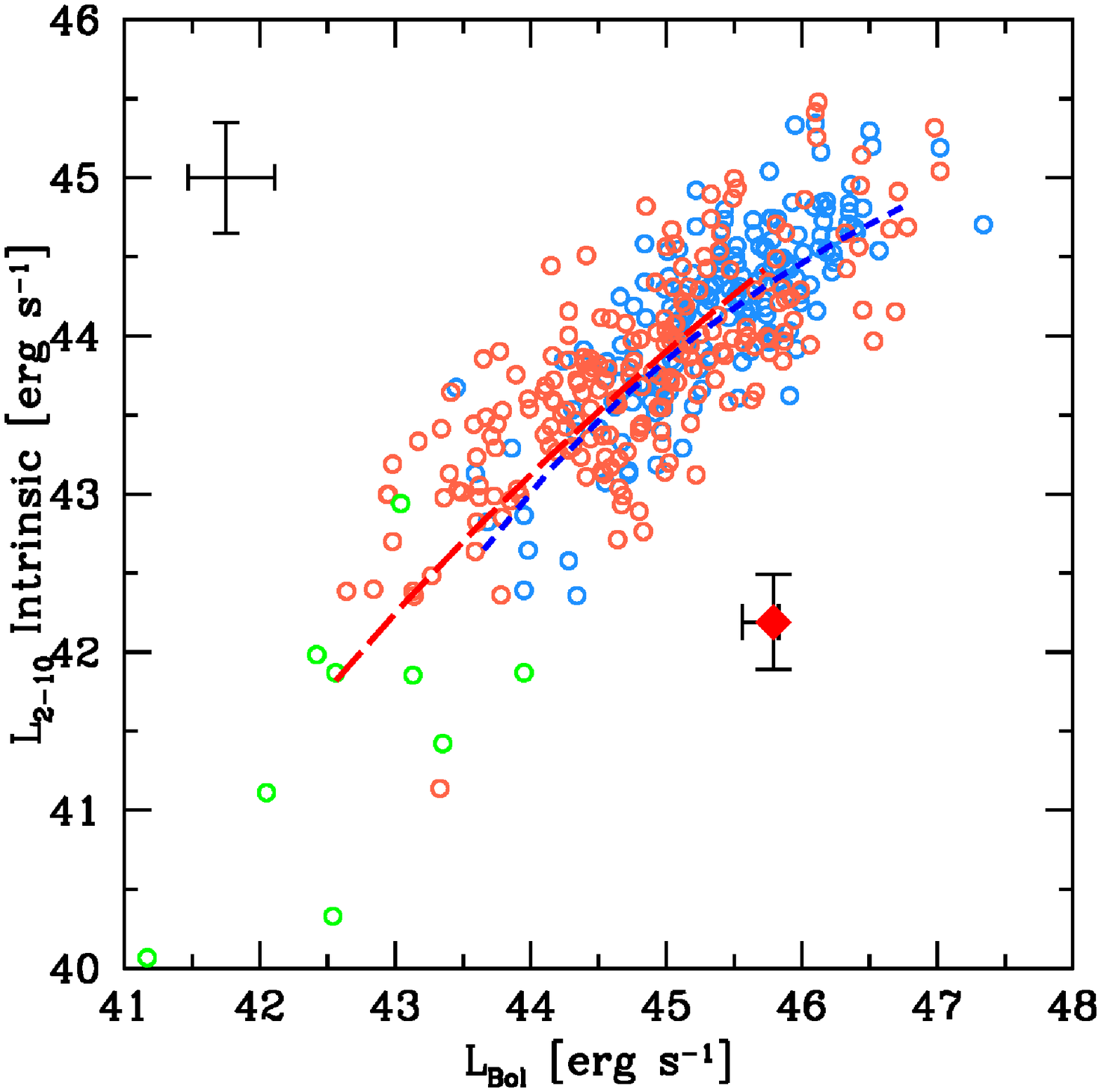}
 \hspace{1.8cm}
 \includegraphics[width=6cm,height=6.5cm]{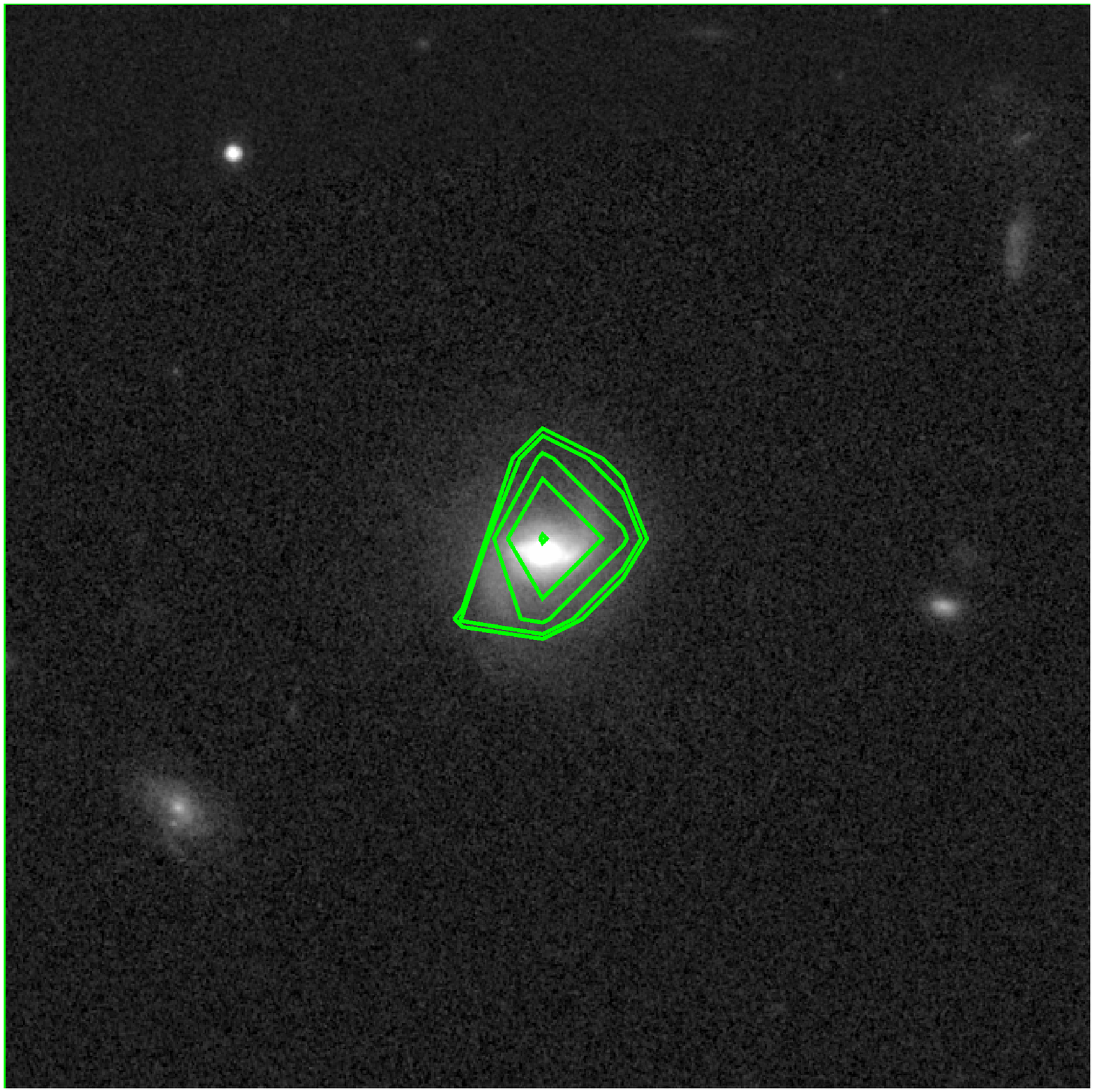}
 \caption{{\it Left:} The intrinsic \lum, corrected for absorption is plotted against the bolometric luminosity derived from the SED fitting. 
 Orange (cyan) circles represent type 2 (type 1) AGN,
while green circles represent galaxies. 
 The average error is shown in the upper left corner.
 Source XID-392 is shown with a red diamond and relative error
bars. 
 The relation derived for type 2 (type 1) AGN in L12 is 
 shown with the red (blue) dashed line. 
 {\it Right:} HST-ACS I band $20\times20\arcsec$ image of the source. The X-ray contours from the \chandra\ full-band (0.5-7 keV) image are superimposed in green. 
 }
 \label{discovery}
 \end{center}
 \end{figure*}
%=======================================================

\section{Selection}

The extreme properties of the source XMMUJ095910.4+020732 in the \xmm\ catalogue (Cappelluti et al. 2007)\footnote{\chandra\ ID: 669, Elvis et al. (2009); 
optical RA=09:59:10.32, DEC=+02:07:32.3, $z=0.353$, Brusa et al. (2010); Civano et al. (2012).}, XID-392 hereafter,
were serendipitously discovered by comparing the intrinsic (absorption-corrected) 
2-10 keV luminosity (\lum) obtained from the automated spectral fit of all the sources in the \xmm\ catalogue (Lanzuisi et al. 2015, L15 hereafter), 
to the bolometric luminosity (\lbol) computed by fitting the spectral energy distribution (SED) for the \textit{\textit{Herschel}-} detected sources in the same field (Delvecchio et al. 2014, 2015).
Figure 1 shows the distribution of the two quantities for all the sample of \xmm\ sources detected by \textit{Herschel}.

The \lbol\ of Fig. 1 (left) represents the total intrinsic luminosity of the accretion disk, derived from the AGN model of the best-fitting SED solution. 
The SED-fitting procedure uses a grid of torus models that were computed by
solving the radiative transfer equation for a smooth dusty structure irradiated by the accretion disc emission (Fritz et al. 2006; Feltre et al. 2012).
This \lbol\ does not include the X-ray emission above $\simgt1$ keV, which is negligible in the total budget, however.
The two quantities plotted in Fig. 1 are related by the X-ray bolometric correction $k_{\rm Bol}$
(\lbol= $k_{\rm Bol}\times$ \lum).
The blue and red curves in Fig. 1 represent the \lbol-dependent $k_{\rm Bol}$ relations derived in Lusso et al. (2012, L12 hereafter) for
type 1 and type 2 AGN in the COSMOS field, respectively.
Strikingly, source XID-392 is the only source in the sample (out of 394) that is more than two orders of magnitudes away from these relations,
and $\sim1.5$ order of magnitudes away from the closest observed point for a given \lum\ or \lbol.

We verified that the source identification is correct and that the IR/optical and X-ray photometry is not contaminated by nearby sources.
Figure 1 (right) shows the HST-ACS  $20\times20\arcsec$ cut-out of the source (Koekemoer et al. 2007). The host galaxy is an isolated barred spiral galaxy seen nearly face-on, and 
the closest source is at a distance of more than $7\arcsec$, which means that the IR/optical photometry is not contaminated. 
The X-ray contours, derived from the \textit{Chandra} full-band (0.5-7 keV) image\footnote{The HST and Chandra cutouts are publicly available at http://irsa.ipac.caltech.edu/data/COSMOS/index\_cutouts.html.}, are superimposed in green.
The off-set between the peak of the X-ray emission and the centre of the host galaxy ($0.1\arcsec$) is well within the \chandra\ absolute astrometric accuracy 
($0.6\arcsec$ of radius for the 90\% uncertainty circle).
Furthermore, XID-392 is the only X-ray emitting source within the $20\times20\arcsec$ area.
Therefore, the remaining possibility is that the absorption-corrected \lum\ is severely underestimated
or that the \lbol\ is severely overestimated.

\section{Multi-wavelength properties}

\subsection{Photometry and SED}

  %======================================================
\begin{figure*}[t]
\begin{center}
\includegraphics[width=7cm,height=5cm]{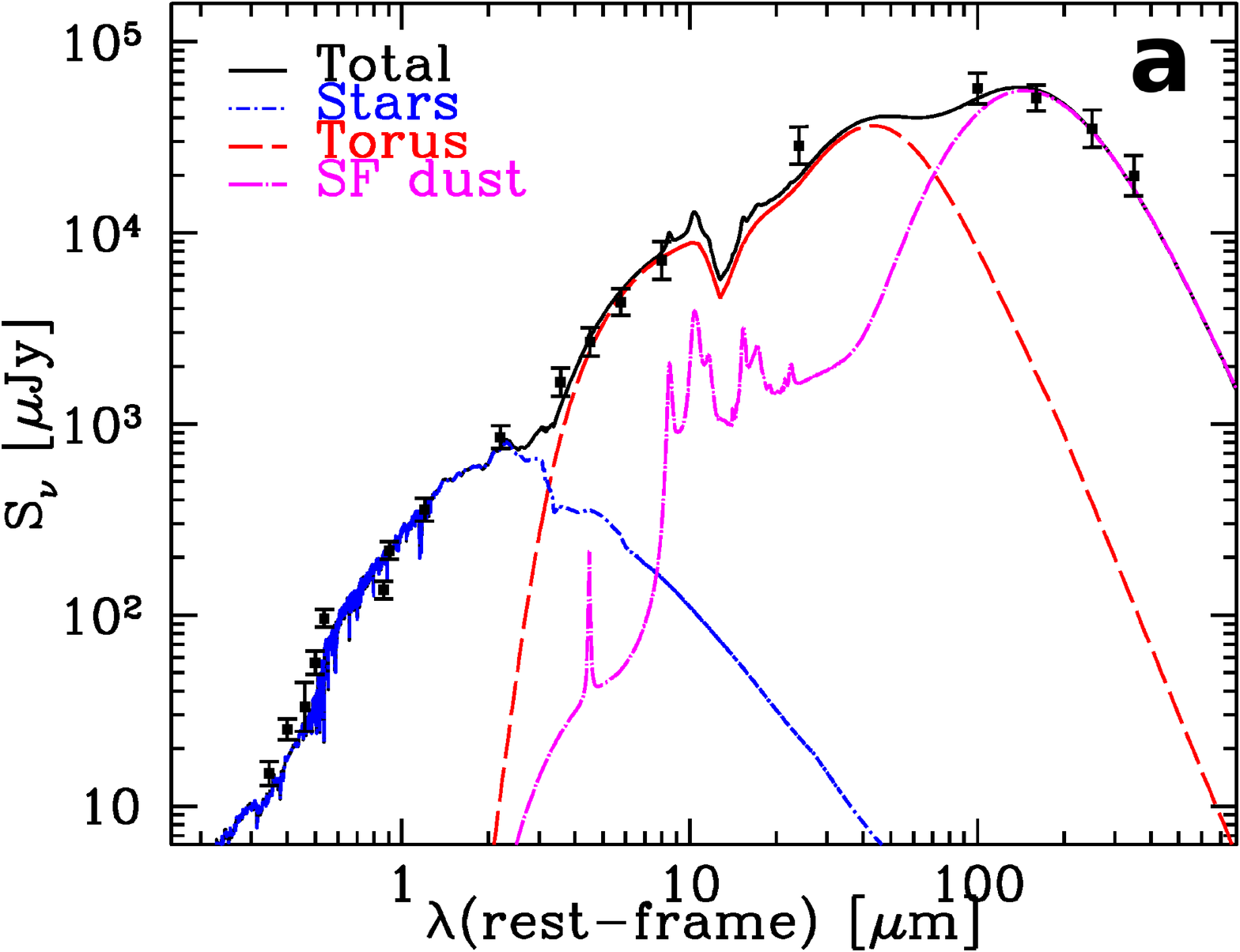}\hspace{1cm}\includegraphics[width=6.5cm,height=5cm]{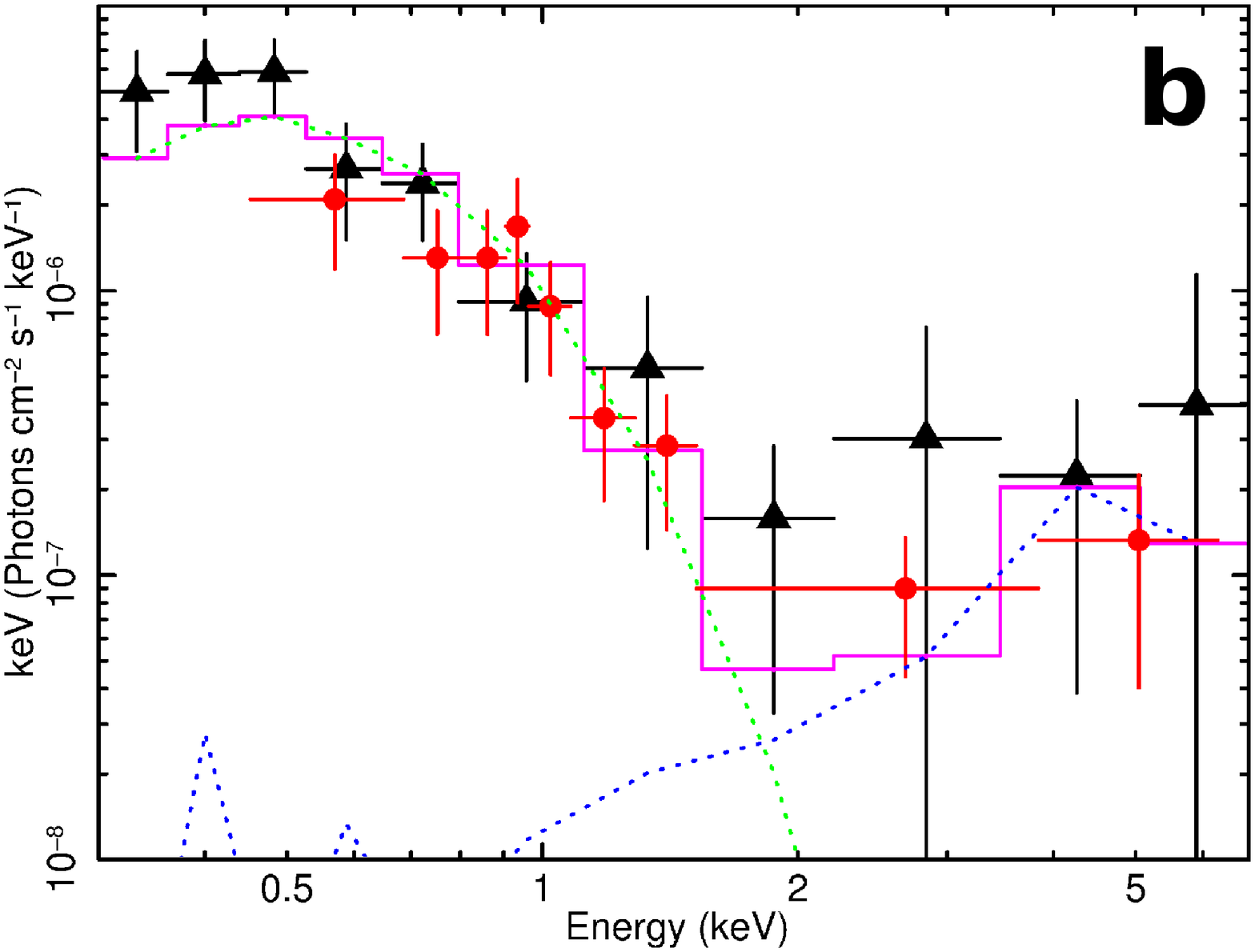}\vspace{0.3cm}
\includegraphics[width=7cm,height=5cm]{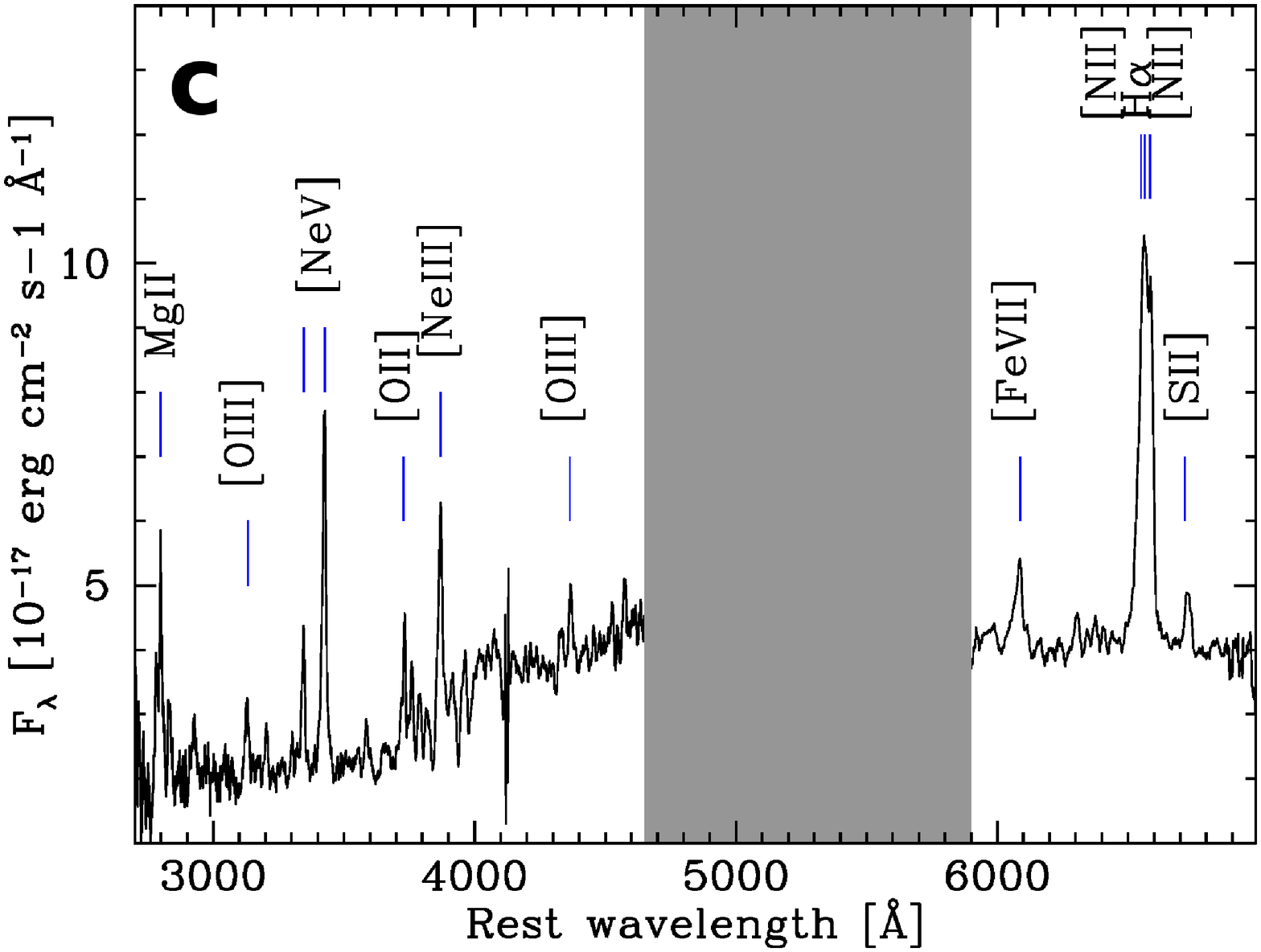}\hspace{0.5cm}\includegraphics[width=9cm,height=5cm]{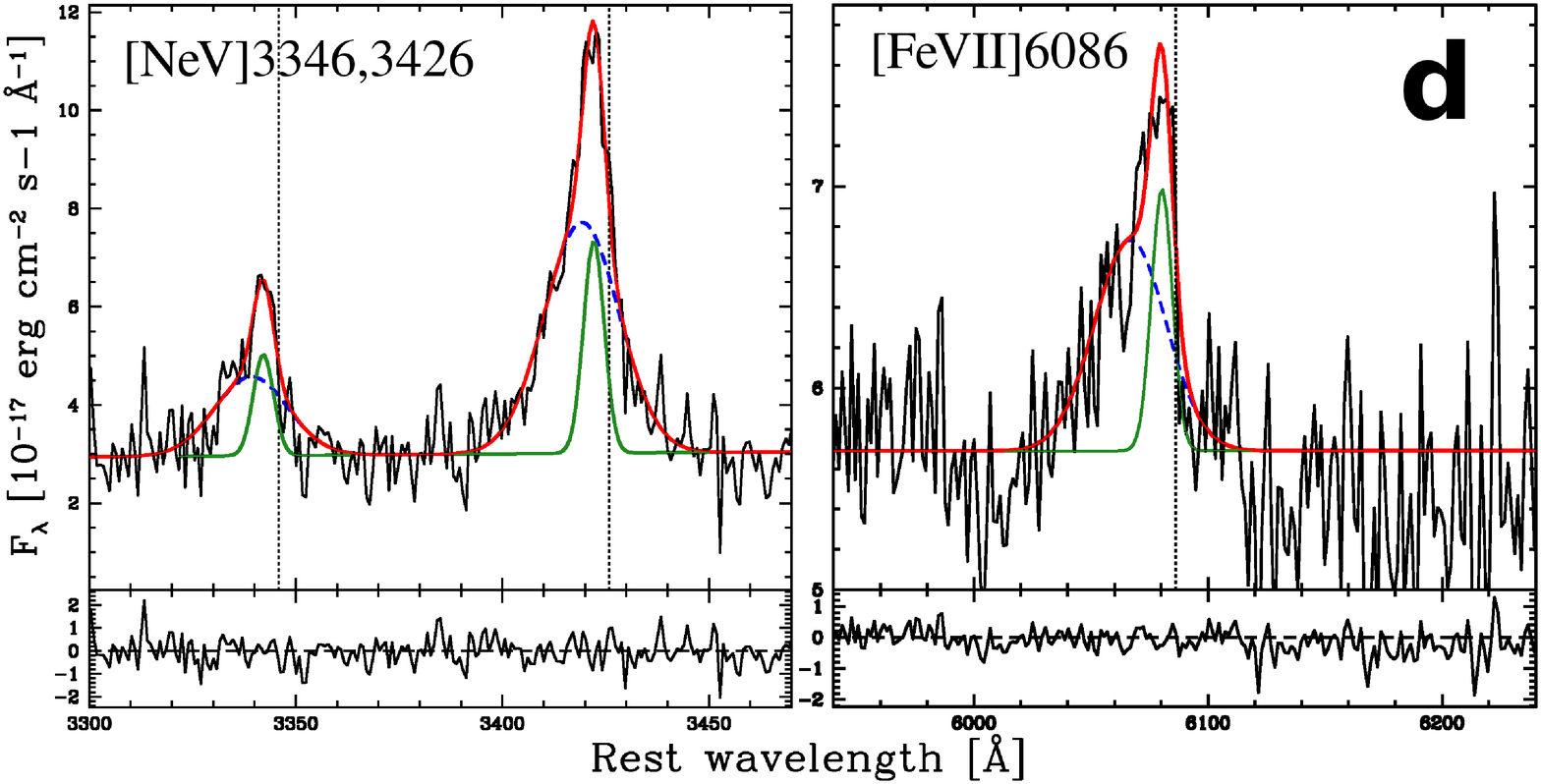}
\caption{{\bf a:} Rest frame, broad-band SED of XID-392. In blue we show the stellar emission, in red the AGN torus emission, and in magenta the SF dust emission.
{\bf b:} Unfolded \xmm\ (black) and \chandra\ (red) spectra of XID-392. 
The best-fit model (magenta) is composed of a thermal component (green) and an obscured torus template (blue). 
{\bf c:} SDSS optical spectrum of XID-392. The most prominent emission lines are labelled. The grey area masks a region of bad-sky subtraction.
{\bf d:} [NeV]3346,3426 and [FeVII]6086 lines decomposition. The green curve represents the narrow component,
while the blue curve represents the broadened asymmetric component, associated with the outflow. The red curve is the sum of the two components. The dotted lines represent the systemic redshift, 
estimated from the continuum and the stellar absorption lines.
}
\label{proper}
\end{center}
\end{figure*}
%======================================================

Figure 2a shows the rest frame broad-band SED fit, performed following Berta et al. (2013).
The source is the third-brightest source at $24\mu m$ in the \xmm\ catalogue
and among the 10\% of the brightest sources in the PACS bands.
The best-fit template includes a prominent AGN torus component (plotted in red) that dominates the source emission 
in all the IRAC (3.6, 4.5, 5.8, and 8 $\mu m$) and MIPS 24 $\mu m$ bands 
(the AGN fraction between 5 and 40 $\mu m$ is 90\%), therefore, the estimate of \lbol\ appears to be robust. 
The galaxy is massive (Log($M_*$)=11.4 \msun), with a derived star formation rate (SFR), 
after subtracting the AGN contribution in the IR, of SFR$=22.2$ \msun\ yr$^{-1}$.
The host galaxy lies at the massive end of the main sequence (MS) of star-forming galaxies (e.g. Whitaker et al. 2012) at z=0.353,
having a specific SFR (sSFR=SFR/$M_*$) of sSFR$=9.3 \times10^{-2}$ Gyr$^{-1}$.

The source is also detected as a compact source at 1.4 GHz in the VLA COSMOS survey (Schinnerer et al. 2004), with a flux of $F_{1.4GHz}=0.424\pm0.025$ mJy. 
Given the definition of the q parameter from Helou et al. (1985, $q=Log(FIR/3.75\times10^{12} Hz)/S_{\nu}(1.4GHz)$), and a far-infrared (FIR, i.e. rest frame 42.5-122.5 $\mu m$) flux of 
FIR$=1.1\times10^{-12}$ \cgs\ of the star-forming component,
XID-392 has $q=1.8\pm0.1$.
Therefore, it has a radio flux that is higher than the average $q$ 
observed in local and high-redshift star-forming galaxies 
($q_{mean} = 2.21\pm0.18$ from Del Moro et al. 2013; $q_{mean} = 2.26\pm0.08$ from Magnelli et al. 2015).
A significant fraction ($\sim45\%$) of these radio-excess galaxies are thought to host low-to-moderate luminosity, or luminous
dust-obscured AGN (Del Moro et al. 2013).

\subsection{X-ray data}

The \chandra\ and \xmm\ spectra of XID-392 were extracted and fitted according to Lanzuisi et al. (2013) and Mainieri et al. (2007).
Figure 2b shows the \chandra\ and \xmm\ spectra with the best-fit model described below.
Both spectra look peculiar, despite the limited photon statistics (64 net counts in \xmm\ and 44 in \chandra).
They show soft emission extending up to 2 keV that can be modelled with either a soft power law with photon index $\Gamma\sim3.7$
or with thermal emission.
The source is not significantly detected in the 2-10 keV band in the \xmm\ data\footnote{The source was not selected as CT in L15 because of the lack of hard detection in the \xmm\ catalogue.}, while the \chandra\ data, 
thanks to the deeper exposure and lower background, enables the detection also in the hard band (with $\sim10$ net counts in the 2-10 keV band). 
Despite the limited spectral quality, the shape of the hard part of the spectrum is clearly flatter
than the typical AGN power law with $\Gamma=1.9$ (Piconcelli et al. 2005): fitting the spectra with a simple power law in the 2-10 keV band alone yields a best-fit photon index of $\Gamma=0.68\pm0.45$.

Assuming that the hard component is produced by obscuration of the primary powerlaw, 
it can be modelled with the torus template from Brightman \& Nandra (2011, consistent results are obtained with mytorus\textup{{\it }} of Murphy \& Yaqoob 2009).
This model gives a best-fit \nh\ value close to $10^{25}$ \cm2.  
Given the limited photon statistics available in the 2-10 keV band, however, this value is poorly constrained.
To estimate at least a lower limit on \nh, we fixed the normalization of the torus{\it } template to 
the normalization required for an intrinsic Log(\lum)=44.25 \ergs, that is, the intrinsic \lum\ expected from \lbol,
using the $k_{\rm Bol}$ of L12.
With this constraint, the resulting lower limit is \nh $\simgt 5\times10^{24}$ \cm2. 
The 2-10 keV observed (not corrected for absorption) luminosity is instead well constrained (Log($L_{2-10}^{Obs}$)=$41.7\pm0.6$ \ergs)\footnote{The \lum\ plotted in Fig. 1 
was computed from the automated fit described in L15
with a single power law and $\Gamma=1.9$ fixed. The observed 2-10 keV luminosity was therefore overestimated and, when considering the spectral model adopted in this work,
the source is even more extreme in terms of distance from the \lum-\lbol\ relation.}.

All nearby CT AGN show a strong ($EW\simgt1$ keV) emission line 
at the rest frame energy of the \feka\ line at 6.4 keV.
This feature is absent from the spectrum of XID-392. 
However, the very limited number of counts available between 6 and 7 keV rest frame ($\text{about}$ four net counts in total)
only allowed us to estimate a loose upper limit for the equivalent width of the \feka\ line of $EW<1.4$(2.4) keV at 90\% ($3\sigma$) confidence level.
The upper limit on the intensity of the \feka\ emission line is therefore fully consistent with the 
possibility XID-392 is a CT AGN.

Given the SFR derived in Sec. 3.1, the expected 0.5-2 and 2-10 keV band luminosities from SFR are L$_{0.5-2}$(SFR)$\sim$L$_{2-10}$(SFR)$\sim10^{41}$ \ergs\  according to the relation reported by Ranalli et al. (2003)
or L$_{0.5-8}$(SFR)$\sim9\times10^{40}$  \ergs from the Mineo et al. (2014) relation.
These values are about one order of magnitude lower than what
was observed from the \chandra\ and \xmm\ spectra of XID-392 
(L(Obs)=1.3, 0.5, 1.6$\times10^{42}$ \ergs in the 0.5-2, 2-10 and 0.5-8 keV band, respectively).
Therefore the presence of a second source of X-ray photons (i.e. the obscured AGN) is required to explain the observed luminosity.

\begin{figure*}[t]
\begin{center}
\includegraphics[width=7cm,height=7cm]{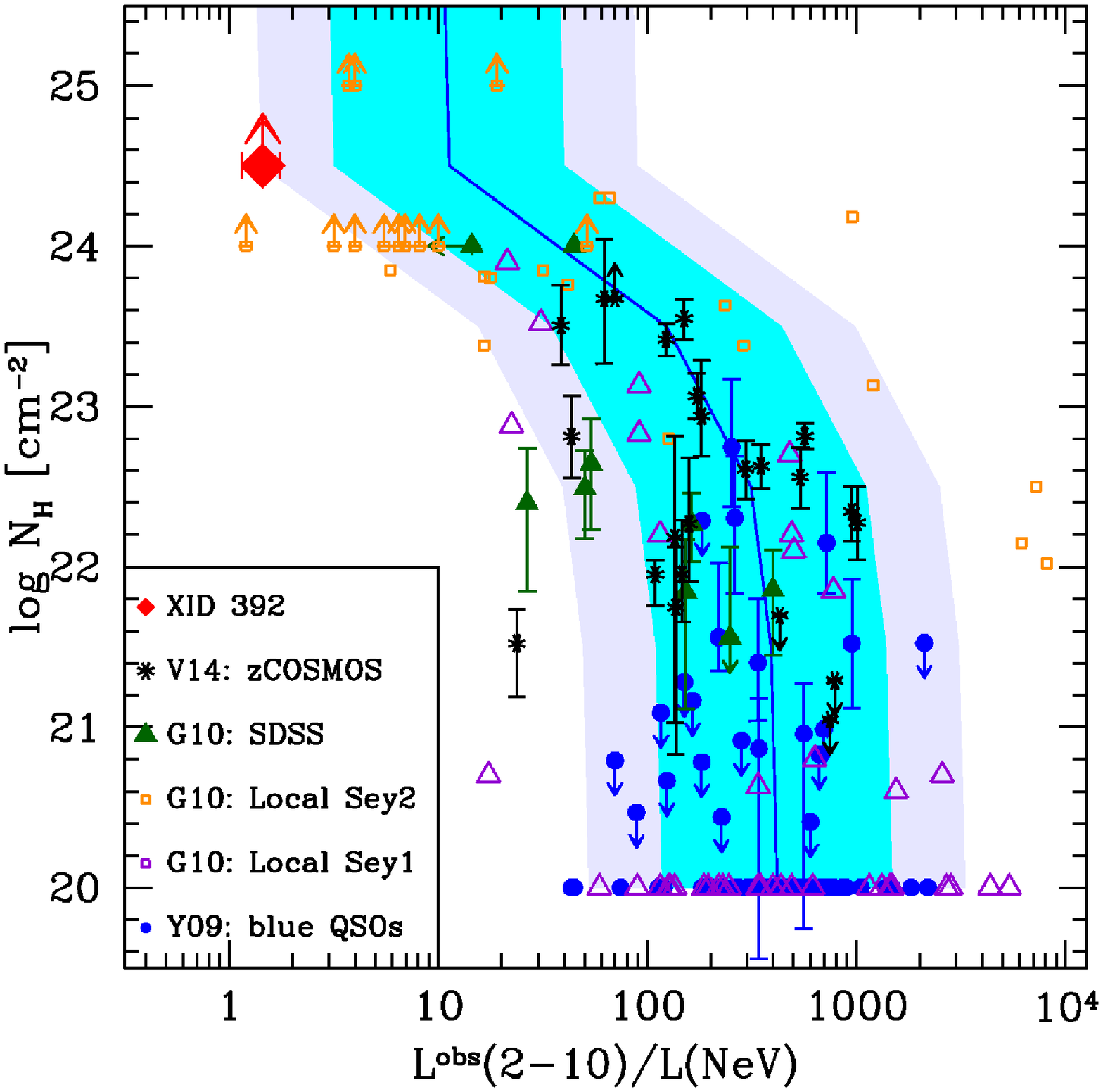}\hspace{1.5cm}
\includegraphics[width=7cm,height=7cm]{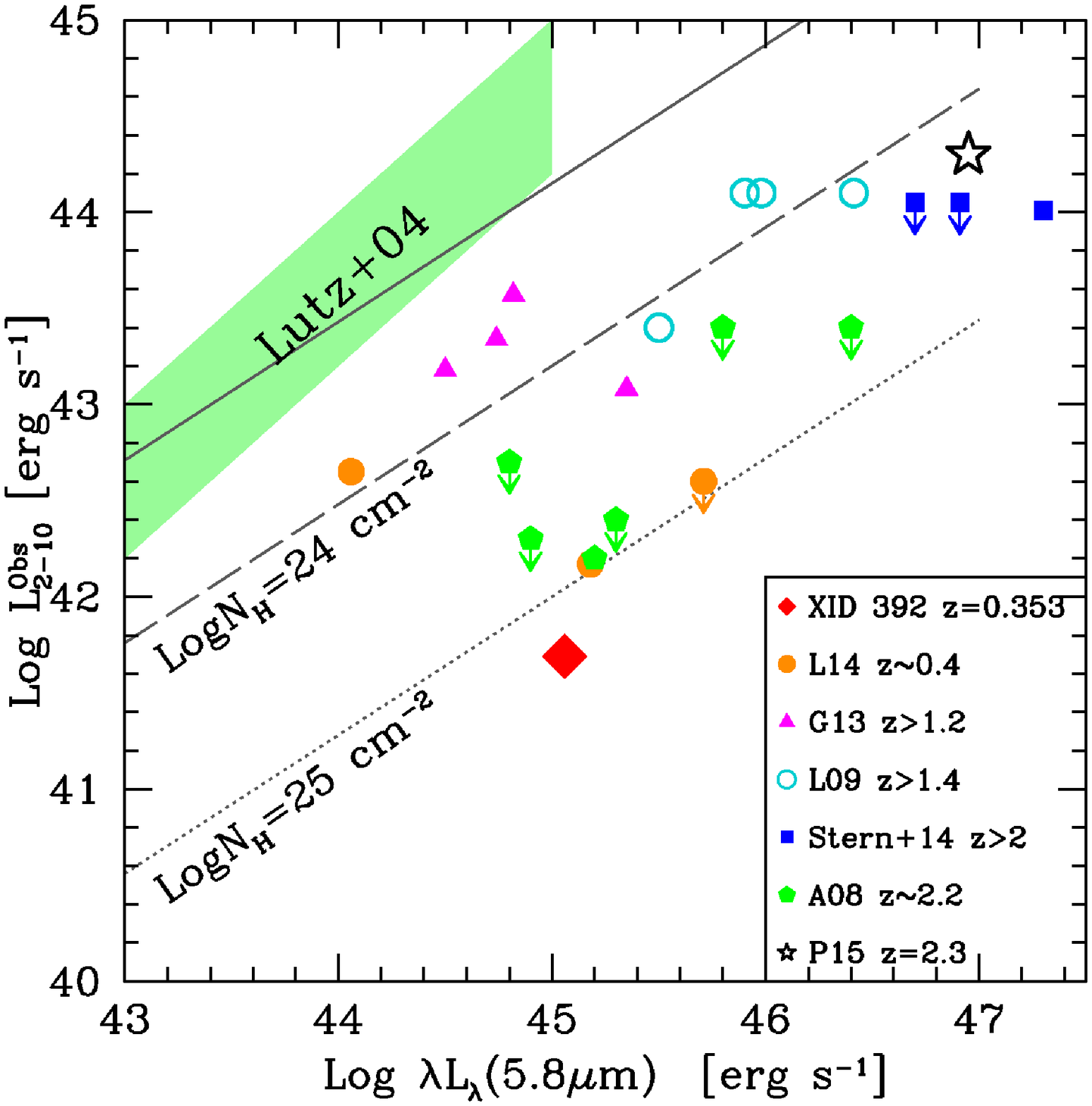}
\caption{{\it Left:} Rest frame 2-10 keV to [NeV] luminosity ratio as a function of \nh. Source XID-392 is shown as a red diamond. V14: Vignali et al. 2014, 
G10: Gilli et al. 2010, Y09: Young et al. 2009. 
{\it Right:} Log$(L_X^{Obs})$ vs Log(\lmir) for several CT candidate (L14: Lansbury et al. 2014, G13: Georgantopoulos et al. 2013, L09: Lanzuisi et al. 2009, 
Stern et al. 2014, A08: Alexander et al. 2008, P15: Piconcelli et al. 2015). 
The green shaded area is the relation reported by Lutz et al. (2004) for a sample of low-redshift AGN.
The black solid line is the relation for the high-redshift AGN described by Fiore et al. (2009), the long dashed (dotted) line the expected relation
for a $10^{24}$ \cm2 ($10^{25}$ \cm2) absorber, starting from the relation given by Fiore et al.}
\label{nh}
\end{center}
\end{figure*}

%======================================================

\section{Obscuration diagnostics}

What we described above shows that the \xray\ data alone, although consistent with 
the presence of an AGN obscured by an HCT absorber, do not provide conclusive evidence.
The rich multi-wavelength data set, both spectroscopic and photometric,
that is available in the COSMOS field, allows us to explore the properties of this unique source in more detail.

\subsection{NeV diagnostics}

Figure 2c shows the SDSS DR10 (Ahn et al. 2013) spectrum of XID-392. 
Several strong emission lines are visible in both the blue and red part of the SDSS spectrum.
Unfortunately, the [OIII] emission line, a commonly used indicator of AGN activity, is masked in the SDSS spectrum because of problems in the sky subtraction.
Although it is typically weaker than [OIII]5007 and suffers stronger dust extinction, the [NeV]3426 
line is considered an unambiguous sign of nuclear activity (Schmidt et al. 1998; Mignoli et al. 2013) because high-energy photons ($> 0.1$ keV) are required to produce this line. 

Gilli et al. (2010) introduced a diagnostic based on the ratio between the observed rest-frame 2-10 keV band and the [NeV]3426 luminosities that was 
calibrated on a sample of obscured and unobscured local Seyferts. 
Figure 3 (left) shows the $L_{2-10}^{obs}/L(NeV)$ ratio for several samples of obscured and unobscured sources (Vignali et al. 2014; Gilli et al. 2010; Young et al. 2009).
The solid line shows the expected trend obtained by starting from the mean ratio $\langle X/[NeV]\rangle$ observed in 
unobscured objects and progressively obscuring the X-ray emission with 
increasing \nh\ (up to log(\nh) = 25.5) while keeping the [NeV] luminosity fixed, because it is produced in the more extended narrow line region. 
The cyan (grey) shaded regions correspond to $\pm1\sigma$ ($\pm90\%$) around the $\langle X/[NeV]\rangle$ ratio (Gilli et al. 2010).
Source XID-392 is shown as a red diamond and has indeed a very strong [NeV] (EW = 43 \AA, F$_{NeV}=1.33\times10^{-15}$ \cgs) and an extreme value of X/[NeV]= $1.44\pm1.20$.
The \nh\ lower limit is derived from the X-ray spectral fit (Sect.~3.2).
Remarkably, XID-392 has the second-lowest X/[NeV] ratio of the sample,
and its value is one order of magnitude lower than the threshold defined in Gilli et al. (2010) to select CT sources.
This confirms the exceptional nature of this source.

\subsection{Mid-IR diagnostics}

The AGN intrinsic \lum\ ($L_X^{Int}$) and the $L_{IR}$ re-emitted by the obscuring torus are known to follow a tight correlation over several orders of magnitudes
(Lutz et al. 2004; Gandhi et al. 2009).
Given that the observed \lum\ ($L_X^{Obs}$) is affected by obscuration, while the $L_{IR}$ is largely independent of it,
the selection of AGN with very low $L_X^{Obs}$ to $L_{IR}$ ratios has been extensively used to identify CT sources.
The $L_X^{Int}$ vs \lmir\ relation is linear for low-redshift Seyfert galaxies, up to \lmir$=10^{45}$ \ergs (Lutz et al. 2004),
while several works on luminous distant QSOs have shown that the relation tends to flatten above \lmir $=10^{44}$ \ergs\ 
(Maiolino et al. 2007; Fiore et al. 2009; Lanzuisi et al. 2009, L09 hereafter).

Figure 3 (right) shows the distribution of $L_X^{Obs}$ vs. \lmir\ for several samples of CT candidates from the CDFN (Alexander et al. 2008),
CDFS (Georgantopoulos et al. 2013), X-SWIRE (L09), and from recent \nus\ observations (Lansbury et al. 2014; Stern et al. 2014).
The range of redshifts covered by the different samples is wide, and even if source XID-392 is at the lowest end of the redshift distribution,
it is intrinsically bright, with \lmir$\sim10^{45}$ \ergs. This
means that it is just at the intersection of the local low-luminosity
and the high-z high-luminosity regime. We stress that the source is the only one below the Log(\nh)=25 \cm2 line (dotted line in Fig. 3 right) 
if the relation described by Fiore et al. (2009) is considered, while it would be even more extreme in terms of $L_X^{Obs}$ vs. \lmir\ for the relation reported by Lutz et al. (2004).

\section{Discussion}

The different diagnostics discussed above strongly indicate that the nuclear emission of XID-392 must be
obscured by an HCT absorber (\nh\ $\sim10^{25}$ \cm2). 
The optical spectrum allow us to explore other physical properties of this unique source. 
An estimate of the SMBH mass (M$_{BH}$) can be obtained from the broad component of the $H\alpha$ line in the optical spectrum
and the 5100\AA\ continuum luminosity or the \lum, using the relations reported in Bongiorno et al. (2014, Eqs. 2 and 4).
 The H$\alpha$-[NII] complex has a total flux of $5.5\times10^{-15}$ \cgs, while the broad component of the H$\alpha$ line 
has a flux of $2.7\times10^{-15}$ \cgs\  and a FWHM of 2700 km/s, with large uncertainties due to the multiple line decompostion.
The 5100\AA\ continuum luminosity is $1.7\times10^{45}$ \ergs.
The resulting M$_{BH}$ are $Log(M_{BH})=8.18$ \msun\ and $Log(M_{BH})=8.35$ \msun\ 
from the two relations, respectively.
Consistent results are obtained if we estimate M$_{BH}$ from the host $M_*$, derived from the SED fitting, 
rescaled to the bulge mass using the mean bulge-to-total mass ratio of 0.2
reported by Kormendy \& Ho (2013) for local barred galaxies. 
This value is further rescaled to the M$_{BH}$ using the BH-to-bulge mass ratio of 0.0023 reported in Marconi \& Hunt (2003)\footnote{The BH-to-bulge mass ratio computed in Kormendy \& Ho (2013) was derived for classical bulges and cannot be 
applied to XID-392, which is a barred spiral.}. The derived value is $Log(M_{BH})\sim8.1$ \msun. 

Given the \lbol\ reported in Sect. 2, the Eddington ratio is \lbol/$L_{Edd}$=0.3-0.5, that is, at the upper end of the distribution for X-ray selected AGN in deep surveys.
The sSFR, close to the MS value, and the face-on barred spiral morphology indicate, however, that XID-392 has not been caught in a major or gas-rich merger event,
but is instead an HCT in which the obscuration is likely to take place in a small-scale torus and not in the host or starburst regions.
Indeed, the presence of a bar in the host is known to correlate with CT obscuration (Maiolino, Risaliti \& Salvati 1999)
and with AGN activity in general (Galloway et al. 2015).

The optical spectrum of XID-392 is complex, showing strong MgII, [NeV], [OII], [NeIII], [NII], and H$\alpha$ emission lines, 
as well as a strong [OIII]3133\AA\ Bowen
fluorescence line (Schachter et al. 1990) and the high-ionization [FeVII]6086\AA\ emission line. 
Furthermore, almost all these emission lines show an asymmetric profile, and remarkably, both the [NeV] and [FeVII], 
having a similarly high ionization potential of $\sim125$ eV.
Figure 2d shows a zoom in the [NeV] and [FeVII] region. The emission lines are fitted with a narrow 
component (green curve) plus a broadened component (blue curve).
Both components are blueshifted with respect to the systemic velocity, 
estimated from the continuum and the stellar absorption lines,
with a typical velocity offset of $\sim300$ and $\sim800$ km/s, respectively. 
The FWHM of the broadened component is in the range 1200-1800 km/s.
These features are commonly associated with outflowing gas on kpc scales 
(Harrison et al. 2012, 2014), and similar features  
have been found in a handful of sources in the XMM-COSMOS catalogue,
with similar accretion properties, that is, high \lbol/$L_{Edd}$ (Brusa et al. 2010; Perna et al. 2015).
The detailed characterization of the properties of the putative outflow are the subject of
ongoing observational effort, and the results will be presented in a forthcoming paper. We stress, however, that 
the exceptional amount of obscuration and the strong outflow signatures are possibly related: 
if the AGN is in the short-lived phase in which powerful accretion (\lbol/$L_{Edd}$=0.3-0.5) 
occurs in a surrounding dense gas cocoon (Hopkins et al. 2005, 2008),
strong outflow episodes, driven by radiation pressure from the AGN, are expected (Menci et al. 2008).

The presence of extreme obscuration and strong outflow signatures
in the same source does not require any particular geometry, except for the fact that the line of sight must
intercept the obscuring material. 
A local example of such a configuration is NGC 1068, which is an HCT  (\nh$>10^{25}$ \cm2\ Matt et al. 2004)
and shows a fast  (v$_{max}=1400$ km/s) outflow with an 
outflow half-opening angle $\theta_{out}=27^{\circ}$ and
an inclination angle of $i=9^{\circ}$ almost perpendicular to the line of sight 
(Muller-Sanchez et al. 2011). 
In a similar scenario that does not imply a wide opening-angle
for the outflow, the blueshift observed in the emission
lines is only the radial component of the real outflow velocity,
and indeed the velocities inferred from the blueshift should
be considered as lower limits.

Detecting one such highly obscured, highly accreting source will not change our 
understanding on the demographics of HCT, of course. 
However, it is important to recognize that these sources do exist beyond the local Universe,
 and they are indeed expected to be much more common at the peak of star formation or BH accretion history,
at $z\sim2-3$.
More importantly, we have demonstrated that they are detectable, 
if a large multi-wavelength survey is combined with medium-deep X-ray observations
and we properly combine all the information simultaneously.
A systematic approach in this direction should be able to retrieve a non-negligible population of such 
HCT sources in a wide range of redshift 
(thanks to the positive K-correction that applies to obscured AGN in X-rays) if applied to a large area, 
multi-wavelength survey such as COSMOS: 
the X-ray background synthesis model described by Gilli et al. (2007) predicts $\sim2.2$ deg$^{-2}$ 
of these HCT sources (Log\nh=25-26 \cm2), 
assuming a flux limit of F$_{2-10}$=$1\times10^{-15}$ \cgs, that
is, the flux observed for a source like XID-392.
Therefore the full COSMOS-Legacy survey (2 deg$^2$, Civano et al. 2015a submitted) will deliver a small but 
valuable sample of such sources. 
We stress that a source like XID-392 is far below the detectability above 10 keV with {\it Swift}-BAT or INTEGRAL-IBIS, 
while \nus\ would require a deep $>500$ ks exposure, which is
not feasible on a large-area survey like COSMOS. The source is not detected in the current $\sim100$ 
ks exposure available in the field (Civano et al. 2015b submitted), which already required a total of 3.2 Ms of \nus\ 
observing time,  therefore the multi-wavelength approach is the only feasible approach for this class of sources with the current
instrumentation.

\begin{acknowledgements}
The authors thank the anonymous referee for useful and constructive comments to the first version of
this paper.
GL, MB, and MP acknowledge financial support from the CIG grant ''eEASY'' n. 321913.
Funding for SDSS-III has been provided by the Alfred P. Sloan Foundation, 
the Participating Institutions, the National Science Foundation, and the U.S. 
Department of Energy Office of Science. The SDSS-III web site is http://www.sdss3.org/.

\end{acknowledgements}

%-------------------------------------------------------------------


\begin{thebibliography}{}


\bibitem[Ahn et al.(2014)]{2014ApJS..211...17A} Ahn, C.~P., Alexandroff, 
R., Allende Prieto, C., et al.\ 2014, \apjs, 211, 17 


\bibitem[Alexander et al.(2008)]{2008ApJ...687..835A} Alexander, D.~M., 
Chary, R.-R., Pope, A., et al.\ 2008, \apj, 687, 835 


\bibitem[Alexander et al.(2013)]{2013ApJ...773..125A} Alexander, D.~M., 
Stern, D., Del Moro, A., et al.\ 2013, \apj, 773, 125 


\bibitem[Berta et 
al.(2013)]{2013A&A...551A.100B} Berta, S., Lutz, D., Santini, P., et al.\ 2013, \aap, 551, AA100 


\bibitem[Bongiorno et al.(2014)]{2014MNRAS.443.2077B} Bongiorno, A., 
Maiolino, R., Brusa, M., et al.\ 2014, \mnras, 443, 2077 




\bibitem[\protect\citeauthoryear{Brightman 
\& Nandra}{2011}]{2011MNRAS.413.1206B} Brightman,  M., \&  Nandra K. 2011, MNRAS, 413, 1206 
 

\bibitem[Brightman et al.(2014)]{2014MNRAS.443.1999B} Brightman, M., 
Nandra, K., Salvato, M., et al.\ 2014, \mnras, 443, 1999 



\bibitem[Brusa et al.(2010)]{2010ApJ...716..348B} Brusa, M., Civano, F., 
Comastri, A., et al.\ 2010, \apj, 716, 348 



\bibitem[Brusa et al.(2015)]{2015MNRAS.446.2394B} Brusa, M., Bongiorno, A., 
Cresci, G., et al.\ 2015, \mnras, 446, 239


\bibitem[Buchner et al.(2015)]{2015arXiv150102805B} Buchner, J., 
Georgakakis, A., Nandra, K., et al.\ 2015, arXiv:1501.02805 


\bibitem[Burlon et al.(2011)]{2011ApJ...728...58B} Burlon, D., Ajello, M., 
Greiner, J., et al.\ 2011, \apj, 728, 58 

\bibitem[Cappelluti et al.(2007)]{2007ApJS..172..341C} Cappelluti, N., 
Hasinger, G., Brusa, M., et al.\ 2007, \apjs, 172, 341 


\bibitem[Chabrier(2003)]{2003ApJ...586L.133C} Chabrier, G.\ 2003, \apjl, 
586, L133 


\bibitem[Civano et al.(2012)]{2012ApJS..201...30C} Civano, F., Elvis, M., 
Brusa, M., et al.\ 2012, \apjs, 201, 30 


\bibitem[Civano et al.(2015)]{2015ApJ submitted}  Civano, F., Marchesi, S., Elvis, M., et al. 2015a, submitted to \apj

\bibitem[Civano et al.(2015)]{2015ApJ submitted}  Civano, F., Hickox, R., Puccetti, S., et al. 2015b, submitted to \apj




\bibitem[Clements et al.(2002)]{2002ApJ...581..974C} Clements, D.~L., 
McDowell, J.~C., Shaked, S., et al.\ 2002, \apj, 581, 974 


\bibitem[Comastri et 
al.(1995)]{1995A&A...296....1C} Comastri, A., Setti, G., Zamorani, G., \& Hasinger, G.\ 1995, \aap, 296, 1 


\bibitem[\protect\citeauthoryear{Comastri et 
al.}{2010}]{2010ApJ...717..787C} Comastri, A., Iwasawa, K., Gilli, R.,  et al.\ 2010, ApJ, 717, 787 

\bibitem[Comastri et 
al.(2015)]{2015A&A...574L..10C} Comastri, A., Gilli, R., Marconi, A., Risaliti, G., \& Salvati, M.\ 2015, \aap, 574, LL10 

\bibitem[Del Moro et 
al.(2013)]{2013A&A...549A..59D} Del Moro, A., Alexander, D.~M., Mullaney, J.~R., et al.\ 2013, \aap, 549, AA59 


\bibitem[Delvecchio et al.(2014)]{2014MNRAS.439.2736D} Delvecchio, I., 
Gruppioni, C., Pozzi, F., et al.\ 2014, \mnras, 439, 2736 


\bibitem[Delvecchio et al.(2015)]{2015arXiv150107602D} Delvecchio, I., 
Lutz, D., Berta, S., et al.\ 2015, arXiv:1501.07602 


\bibitem[Draper 
\& Ballantyne(2012)]{2012ApJ...753L..37D} Draper, A.~R., \& Ballantyne, D.~R.\ 2012, \apjl, 753, LL37 


\bibitem[Elvis et al.(2009)]{2009ApJS..184..158E} Elvis, M., Civano, F., 
Vignali, C., et al.\ 2009, \apjs, 184, 158 

\bibitem[Feltre et al.(2012)]{2012MNRAS.426..120F} Feltre, A., 
Hatziminaoglou, E., Fritz, J., \& Franceschini, A.\ 2012, \mnras, 426, 120 



\bibitem[Fiore et al.(2009)]{2009ApJ...693..447F} Fiore, F., Puccetti, S., 
Brusa, M., et al.\ 2009, \apj, 693, 447 


\bibitem[Fritz et al. (2006)]{fritz2006} 
Fritz, J., Franceschini, A., \& Hatziminaoglou, E.\ 2006, \mnras, 366, 767 


\bibitem[Galloway et al.(2015)]{2015arXiv150201033G} Galloway, M.~A., 
Willett, K.~W., Fortson, L.~F., et al.\ 2015, arXiv:1502.01033 


\bibitem[Gandhi et 
al.(2009)]{2009A&A...502..457G} Gandhi, P., Horst, H., Smette, A., et al.\ 2009, \aap, 502, 457 


\bibitem[Genzel et al.(1998)]{1998ApJ...498..579G} Genzel, R., Lutz, D., 
Sturm, E., et al.\ 1998, \apj, 498, 579 


\bibitem[Georgantopoulos et 
al.(2013)]{2013A&A...555A..43G} Georgantopoulos, I., Comastri, A., Vignali, C., et al.\ 2013, \aap, 555, AA43 



\bibitem[Gilli et 
al.(2007)]{2007A&A...463...79G} Gilli, R., Comastri, A., \& Hasinger, G.\ 2007, \aap, 463, 79 


\bibitem[Gilli et 
al.(2010)]{2010A&A...519A..92G} Gilli, R., Vignali, C., Mignoli, M., et al.\ 2010, \aap, 519, AA92 


\bibitem[Graham 
\& Scott(2013)]{2013ApJ...764..151G} Graham, A.~W., \& Scott, N.\ 2013, \apj, 764, 151 



\bibitem[Harrison et al.(2012)]{2012MNRAS.426.1073H} Harrison, C.~M., 
Alexander, D.~M., Swinbank, A.~M., et al.\ 2012, \mnras, 426, 1073 

\bibitem[Harrison et al.(2014)]{2014MNRAS.441.3306H} Harrison, C.~M., 
Alexander, D.~M., Mullaney, J.~R., 
\& Swinbank, A.~M.\ 2014, \mnras, 441, 3306 



\bibitem[Harrison et al.(2013)]{2013ApJ...770..103H} Harrison, F.~A., 
Craig, W.~W., Christensen, F.~E., et al.\ 2013, \apj, 770, 103 


\bibitem[Helou et al.(1985)]{1985ApJ...298L...7H} Helou, G., Soifer, B.~T., 
\& Rowan-Robinson, M.\ 1985, \apjl, 298, L7 


\bibitem[Hopkins et al.(2005)]{2005ApJ...625L..71H} Hopkins, P.~F., 
Hernquist, L., Martini, P., et al.\ 2005, \apjl, 625, L71 



\bibitem[Hopkins et al.(2008)]{2008ApJS..175..356H} Hopkins, P.~F., 
Hernquist, L., Cox, T.~J., \& Kere{\v s}, D.\ 2008, \apjs, 175, 356 


\bibitem[Iwasawa et al.(2001)]{2001MNRAS.326..894I} Iwasawa, K., Matt, G., 
Guainazzi, M., \& Fabian, A.~C.\ 2001, \mnras, 326, 894 


\bibitem[Koekemoer et al.(2007)]{2007ApJS..172..196K} Koekemoer, A.~M., 
Aussel, H., Calzetti, D., et al.\ 2007, \apjs, 172, 196 


\bibitem[Kormendy 
\& Ho(2013)]{2013ARA&A..51..511K} Kormendy, J., \& Ho, L.~C.\ 2013, \araa, 51, 511 

\bibitem[Lansbury et al.(2014)]{2014ApJ...785...17L} Lansbury, G.~B., 
Alexander, D.~M., Del Moro, A., et al.\ 2014, \apj, 785, 17 



\bibitem[Lanzuisi et 
al.(2009)]{2009A&A...498...67L} Lanzuisi, G., Piconcelli, E., Fiore, F., et al.\ 2009, \aap, 498, 67 (L09)

\bibitem[Lanzuisi et al.(2013)]{2013MNRAS.431..978L} Lanzuisi, G., Civano, 
F., Elvis, M., et al.\ 2013, \mnras, 431, 978 

\bibitem[Lanzuisi et 
al.(2015)]{2015A&A...573A.137L} Lanzuisi, G., Ranalli, P., Georgantopoulos, I., et al.\ 2015, \aap, 573, AA137 (L15)


\bibitem[Lusso et al.(2012)]{2012MNRAS.425..623L} Lusso, E., Comastri, A., 
Simmons, B.~D., et al.\ 2012, \mnras, 425, 623 (L12)

\bibitem[Lutz et 
al.(2004)]{2004A&A...418..465L} Lutz, D., Maiolino, R., Spoon, H.~W.~W., \& Moorwood, A.~F.~M.\ 2004, \aap, 418, 465 


\bibitem[Magnelli et 
al.(2015)]{2015A&A...573A..45M} Magnelli, B., Ivison, R.~J., Lutz, D., et al.\ 2015, \aap, 573, AA45 


\bibitem[Mainieri et al.(2007)]{2007ApJS..172..368M} Mainieri, V., 
Hasinger, G., Cappelluti, N., et al.\ 2007, \apjs, 172, 368 



\bibitem[Maiolino et 
al.(1998)]{1998A&A...338..781M} Maiolino, R., Salvati, M., Bassani, L., et al.\ 1998, \aap, 338, 781 



\bibitem[Maiolino et 
al.(1999)]{1999A&A...341L..35M} Maiolino, R., Risaliti, G., \& Salvati, M.\ 1999, \aap, 341, L35 

\bibitem[Maiolino et 
al.(2007)]{2007A&A...468..979M} Maiolino, R., Shemmer, O., Imanishi, M., et al.\ 2007, \aap, 468, 979 

\bibitem[Marconi 
\& Hunt(2003)]{2003ApJ...589L..21M} Marconi, A., \& Hunt, L.~K.\ 2003, \apjl, 589, L21 



\bibitem[Menci et al.(2008)]{2008ApJ...686..219M} Menci, N., Fiore, F., 
Puccetti, S., \& Cavaliere, A.\ 2008, \apj, 686, 219 



\bibitem[Mignoli et 
al.(2013)]{2013A&A...556A..29M} Mignoli, M., Vignali, C., Gilli, R., et al.\ 2013, \aap, 556, AA29 


\bibitem[Mineo et al.(2014)]{2014MNRAS.437.1698M} Mineo, S., Gilfanov, M., 
Lehmer, B.~D., Morrison, G.~E., \& Sunyaev, R.\ 2014, \mnras, 437, 1698 

\bibitem[M{\"u}ller-S{\'a}nchez et al.(2011)]{2011ApJ...739...69M} 
M{\"u}ller-S{\'a}nchez, F., Prieto, M.~A., Hicks, E.~K.~S., et al.\ 2011, 
\apj, 739, 69 

\bibitem[Murphy 
\& Yaqoob(2009)]{2009MNRAS.397.1549M} Murphy, K.~D., \& Yaqoob, T.\ 2009, \mnras, 397, 1549 


\bibitem[Nardini et al.(2010)]{2010MNRAS.405.2505N} Nardini, E., Risaliti, 
G., Watabe, Y., Salvati, M., \& Sani, E.\ 2010, \mnras, 405, 2505 


\bibitem[Perna et 
al.(2015)]{2015A&A...574A..82P} Perna, M., Brusa, M., Cresci, G., et al.\ 2015, \aap, 574, AA82 



\bibitem[Piconcelli et 
al.(2005)]{2005A&A...432...15P} Piconcelli, E., Jimenez-Bail{\'o}n, E., Guainazzi, M., et al.\ 2005, \aap, 432, 15

\bibitem[Piconcelli et 
al.(2015)]{2015A&A...574L...9P} Piconcelli, E., Vignali, C., Bianchi, S., et al.\ 2015, \aap, 574, LL9 



\bibitem[Ranalli et  al.(2003)]{2003A&A...399...39R} Ranalli, P., Comastri, A., \& Setti, G.\ 2003, \aap, 399, 39 


\bibitem[Risaliti et al.(1999)]{1999ApJ...522..157R} Risaliti, G., 
Maiolino, R., \& Salvati, M.\ 1999, \apj, 522, 157 


\bibitem[Schachter et al.(1990)]{1990ApJ...362...74S} Schachter, J., 
Filippenko, A.~V., \& Kahn, S.~M.\ 1990, \apj, 362, 74 


\bibitem[Schinnerer et al.(2004)]{2004AJ....128.1974S} Schinnerer, E., 
Carilli, C.~L., Scoville, N.~Z., et al.\ 2004, \aj, 128, 1974 

\bibitem[Schmitt(1998)]{1998ApJ...506..647S} Schmitt, H.~R.\ 1998, \apj, 
506, 647 


\bibitem[Scoville et al.(2007)]{2007ApJS..172....1S} Scoville, N., Aussel, 
H., Brusa, M., et al.\ 2007, \apjs, 172, 1 

\bibitem[Scoville et al.(2015)]{2015ApJ...800...70S} Scoville, N., Sheth, 
K., Walter, F., et al.\ 2015, \apj, 800, 70 


\bibitem[Stern et al.(2014)]{2014ApJ...794..102S} Stern, D., Lansbury, 
G.~B., Assef, R.~J., et al.\ 2014, \apj, 794, 102 



\bibitem[Tozzi et 
al.(2006)]{2006A&A...451..457T} Tozzi, P., Gilli, R., Mainieri, V., et al.\ 2006, \aap, 451, 457 

\bibitem[Ueda et al.(2014)]{2014ApJ...786..104U} Ueda, Y., Akiyama, M., 
Hasinger, G., Miyaji, T., \& Watson, M.~G.\ 2014, \apj, 786, 104 


\bibitem[Vignali et 
al.(2014)]{2014A&A...571A..34V} Vignali, C., Mignoli, M., Gilli, R., et al.\ 2014, \aap, 571, AA34 


\bibitem[Wilson et al.(2014)]{2014ApJ...789L..36W} Wilson, C.~D., Rangwala, 
N., Glenn, J., et al.\ 2014, \apjl, 789, LL36 


\bibitem[Whitaker et al.(2012)]{2012ApJ...754L..29W} Whitaker, K.~E., van 
Dokkum, P.~G., Brammer, G., \& Franx, M.\ 2012, \apjl, 754, LL29 


\bibitem[Young et al.(2009)]{2009ApJS..183...17Y} Young, M., Elvis, M., 
\& Risaliti, G.\ 2009, \apjs, 183, 17 




\end{thebibliography}
\end{document}